\documentclass[journal,twoside]{IEEEtran}
\usepackage{amsmath}
\usepackage{amssymb}
\usepackage{amsfonts}
\usepackage{bm}
\usepackage{mathtools}
\ifCLASSINFOpdf
\else
\fi

\usepackage{amsmath,amsfonts}
\usepackage{algorithmic}
\usepackage{algorithm}
\usepackage{array}
\usepackage[caption=false,font=normalsize,labelfont=sf,textfont=sf]{subfig}
\usepackage{textcomp}
\usepackage{stfloats}
\usepackage{url}
\usepackage{verbatim}
\usepackage{graphicx}
\usepackage{cite}

\usepackage{booktabs}
\usepackage{siunitx}
\usepackage{xcolor}
\usepackage{enumitem} 
\usepackage[expansion=false]{microtype} 
\usepackage{tabularx}
\usepackage{multirow}
\usepackage{tikz} 
\usetikzlibrary{shapes.geometric, arrows.meta, positioning, calc, shadows.blur, backgrounds, fit, patterns}
\usepackage{pgfplots} 
\pgfplotsset{compat=1.18}
\usepackage{placeins} 
\usepackage[colorlinks=true, allcolors=blue]{hyperref} 

\definecolor{myblue}{HTML}{5499C7}
\definecolor{myorange}{HTML}{F39C12}
\definecolor{mygreen}{HTML}{52BE80}
\definecolor{myred}{HTML}{E74C3C}
\definecolor{mygray}{HTML}{EAECEE}
\definecolor{mypurple}{HTML}{8E44AD}
\definecolor{datagen_blue}{HTML}{2980B9}
\definecolor{datagen_green}{HTML}{27AE60}
\definecolor{datagen_red}{HTML}{C0392B}
\definecolor{datagen_gray}{HTML}{ECF0F1}
\definecolor{hist_bar_blue}{HTML}{3498DB}
\definecolor{hist_bar_red}{HTML}{E74C3C}

\tikzstyle{io} = [trapezium, trapezium left angle=70, trapezium right angle=110, minimum width=3.2cm, minimum height=0.8cm, text centered, draw=black, fill=myblue!80, text=white, rounded corners=3pt, blur shadow={shadow xshift=0.5ex, shadow yshift=-0.5ex}, font=\sffamily\footnotesize, text width=3cm]
\tikzstyle{process} = [rectangle, minimum width=3.2cm, minimum height=0.8cm, text centered, text width=3.2cm, draw=black, fill=myorange!80, text=white, rounded corners=3pt, blur shadow={shadow xshift=0.5ex, shadow yshift=-0.5ex}, font=\sffamily\footnotesize]
\tikzstyle{decision} = [diamond, aspect=1.8, minimum width=2.5cm, minimum height=0.8cm, text centered, draw=black, fill=myred!80, text=white, rounded corners=3pt, blur shadow={shadow xshift=0.5ex, shadow yshift=-0.5ex}, font=\sffamily\footnotesize, text width=2.5cm]
\tikzstyle{output_node} = [rectangle, minimum width=3.2cm, minimum height=0.8cm, text centered, text width=3.2cm, draw=black, fill=mygreen!80, text=white, rounded corners=3pt, blur shadow={shadow xshift=0.5ex, shadow yshift=-0.5ex}, font=\sffamily\footnotesize]
\tikzstyle{solver} = [rectangle, minimum width=3.2cm, minimum height=0.8cm, text centered, text width=3.2cm, draw=black, fill=mypurple!80, text=white, rounded corners=3pt, blur shadow={shadow xshift=0.5ex, shadow yshift=-0.5ex}, font=\sffamily\footnotesize]
\tikzstyle{arrow} = [thick, -{Stealth}, rounded corners=3pt]

\tikzstyle{gnn_input} = [rectangle, rounded corners=3pt, minimum width=3.2cm, minimum height=0.9cm, text centered, draw=black, fill=myblue!80, text=white, align=center, blur shadow={shadow xshift=0.5ex, shadow yshift=-0.5ex}, font=\sffamily\footnotesize, text width=3cm]
\tikzstyle{gnn_layer} = [rectangle, rounded corners=3pt, minimum width=3.2cm, minimum height=1.2cm, text centered, draw=black, fill=myorange!80, text=white, align=center, blur shadow={shadow xshift=0.5ex, shadow yshift=-0.5ex}, font=\sffamily\footnotesize, text width=3cm]
\tikzstyle{gnn_output} = [rectangle, rounded corners=3pt, minimum width=3.2cm, minimum height=0.9cm, text centered, draw=black, fill=mygreen!80, text=white, align=center, blur shadow={shadow xshift=0.5ex, shadow yshift=-0.5ex}, font=\sffamily\footnotesize, text width=3cm]
\tikzstyle{gnn_arrow} = [thick,-{Stealth},rounded corners=3pt]
\tikzstyle{gnn_resid} = [circle, draw=black, fill=white, inner sep=0pt, minimum size=5mm, font=\sffamily\bfseries, blur shadow={shadow xshift=0.5ex, shadow yshift=-0.5ex}]

\tikzstyle{datagen_startstop} = [rectangle, rounded corners, minimum width=3cm, minimum height=0.8cm, text centered, draw=black, fill=datagen_gray, font=\sffamily\footnotesize]
\tikzstyle{datagen_process} = [rectangle, minimum width=3cm, minimum height=0.8cm, text centered, text width=3cm, draw=black, fill=datagen_blue!80, text=white, rounded corners=3pt, font=\sffamily\footnotesize]
\tikzstyle{datagen_decision} = [diamond, aspect=1.8, minimum width=2.5cm, text centered, draw=black, fill=datagen_red!80, text=white, rounded corners=3pt, font=\sffamily\footnotesize, text width=2.5cm]
\tikzstyle{datagen_io} = [trapezium, trapezium left angle=70, trapezium right angle=110, minimum width=3cm, minimum height=0.8cm, text centered, draw=black, fill=myorange!80, text=white, rounded corners=3pt, font=\sffamily\footnotesize]
\tikzstyle{datagen_save} = [rectangle, minimum width=3cm, minimum height=0.8cm, text centered, text width=3cm, draw=black, fill=datagen_green!80, text=white, rounded corners=3pt, font=\sffamily\footnotesize]
\tikzstyle{datagen_arrow} = [thick, -{Stealth}, rounded corners=3pt]

\newcommand{\thead}[1]{\textbf{\begin{tabular}{@{}c@{}}#1\end{tabular}}}

\pgfplotsset{every axis/.append style={
		tick label style={font=\footnotesize},
		label style={font=\footnotesize},
		title style={font=\footnotesize},
		legend style={font=\footnotesize}
}}

\AtBeginEnvironment{figure}{\centering}

\begin{document}
	
	\title{A Hybrid GNN-IZR Framework for Fast and Empirically Robust AC Power Flow Analysis in Radial Distribution Systems}
	
	\author{Mohamed Shamseldein,~\IEEEmembership{Senior Member,~IEEE}
		\thanks{M. Shamseldein is with the Department of Electrical Power and Machines, Faculty of Engineering, Ain Shams University, Cairo, Egypt (e-mail: mohamed.shamseldein@eng.asu.edu.eg).}%
		\thanks{Manuscript received September 14, 2025; revised ...}}
	
	\markboth{IEEE Transactions on Power Systems,~Vol.~XX, No.~Y, September~2025}%
	{Shamseldein: A Hybrid GNN-IZR Framework for Fast ACPF}
	
	
	\maketitle
	
	\begin{abstract}
		The Alternating Current Power Flow (ACPF) problem forces a trade-off between the speed of data-driven models and the reliability of analytical solvers. This paper introduces a hybrid framework that synergizes a Graph Neural Network (GNN) with the Implicit Z-Bus Recursive (IZR) method, a robust, non-iterative solver for radial distribution networks. The framework employs a physics-informed GNN for rapid initial predictions and invokes the IZR solver as a failsafe for stressed cases identified by a two-stage trigger. A failure is defined as any solution with a maximum power mismatch exceeding \SI{0.1}{p.u.}, a significant operational deviation. On a challenging test set of 7,500 stressed scenarios for the IEEE 33-bus system, the GNN-only model failed on 13.11\% of cases. In contrast, the hybrid framework identified all potential failures, delegating them to the IZR solver to achieve a 0.00\% failure rate, empirically matching the 100\% success rate of the analytical solver on this specific test set. An expanded ablation study confirms that both physics-informed training and Z-bus sensitivity features are critical, collaboratively reducing the GNN's failure rate from 98.72\% (data-only) to 13.11\%. The hybrid approach demonstrates a pragmatic path to achieving the empirical reliability of an analytical solver while leveraging GNN speed, enabling a significant increase in the number of scenarios analyzable in near real-time.
	\end{abstract}
	
	
	\section{Introduction}
	\IEEEPARstart{T}{he} increasing penetration of variable renewable energy sources is transforming the Alternating Current Power Flow (ACPF) problem from a routine planning task into a critical, real-time operational imperative. This creates a fundamental "speed-reliability dilemma": the need for faster calculations to ensure dynamic security clashes with the non-negotiable requirement for solver reliability in critical infrastructure \cite{ref_wood_wollenberg}. 
	These challenges are particularly acute in radial distribution systems, where the massive integration of distributed energy resources (DERs) creates unprecedented operational complexity, motivating the need for specialized, high-speed analytical tools.
	Reducing solution time from tens of milliseconds to under four milliseconds can enable operators to analyze thousands more contingency scenarios per minute, drastically improving situational awareness.
	
	Traditional solvers like Newton-Raphson (NR) are robust for well-behaved systems but can struggle with convergence in ill-conditioned cases, such as the heavily stressed scenarios explored in this study \cite{ref_stott_review}. Conversely, data-driven methods, particularly Graph Neural Networks (GNNs), offer near-instantaneous solutions but often have non-negligible failure rates under stressed conditions, a major barrier to their adoption \cite{ref_khaloie_review, ref_karniadakis_review, ref_yaniv_adoption}. In this context, a "failure" is defined as any solution that violates physical laws, quantified as a maximum power mismatch at any bus exceeding \SI{0.1}{p.u.}.
	This threshold was selected as it represents a significant, operationally unacceptable deviation from the specified power injection, indicating a physically implausible state.
	
	This paper introduces a Hybrid GNN-IZR framework that strategically resolves this dilemma by combining the strengths of both approaches: the GNN's speed for nominal cases and the IZR solver's reliability for difficult ones.
	The main contributions are: 1) a hybrid framework leveraging a GNN for speed and the IZR analytical solver for robustness, which empirically reduces the system's prediction failure rate from 13.11\% (GNN-only) to zero on the test set; 2) an expanded ablation study that disentangles the effects of the physics-informed loss and Z-bus features, demonstrating both are critical for reducing the GNN's failure rate; 3) a formal analysis of the framework's two-stage robustness trigger, quantifying its performance as a binary classifier for detecting potential GNN failures; 4) a detailed characterization of the framework's bimodal runtime performance using percentile statistics and histograms; and 5) a stabilized training strategy incorporating dynamic loss scaling, three-phase physics loss annealing, and a Pareto-inspired checkpointing criterion to ensure robust model convergence.
	
	\section{Background and Related Work}
	
	\subsection{Analytical Power Flow Solvers}
	The Newton-Raphson (NR) method is the industry standard for ACPF \cite{ref_stagg_elabiad}. However, its performance can degrade in ill-conditioned systems, such as stressed radial distribution networks \cite{ref_iwamoto_tamura}. 
	
	The IZR method offers a robust, non-iterative alternative for radial and weakly-meshed networks \cite{ref_shamseldein_izr}. It models bus voltages as holomorphic functions of a complex parameter, $\alpha$, that scales the constant power injections. These functions are then expanded using a Maclaurin series. The core power flow equation for all non-slack buses is embedded with $\alpha$:
	\begin{equation}
		\tilde{\mathbf{V}}(\alpha) = \left(\tilde{\mathbf{Y}}\right)^{-1}(\operatorname{diag}(\alpha\tilde{\mathbf{S}}^{*})\tilde{\mathbf{W}}^{*}(\alpha^{*}) + \tilde{\mathbf{I}}_{L} - \mathbf{y}V_{1})
	\end{equation}
	where $\tilde{\mathbf{Y}}$ is the admittance matrix for non-slack buses, $\tilde{\mathbf{S}}$ and $\tilde{\mathbf{I}}_{L}$ are the constant power and current injections, and $\tilde{\mathbf{W}}^{*}(\alpha^{*})$ is the conjugate of the reciprocal voltage vector. The $(\tilde{\mathbf{Y}})^{-1}$ term represents the implicit Z-bus matrix. The tilde notation indicates submatrices corresponding to non-slack buses, and the conjugation and reciprocal operations are component-wise.
	
	The solution begins at $\alpha=0$, representing a system with only constant impedance and current loads, which yields the initial voltage vector $\tilde{\mathbf{V}}[0]$:
	\begin{equation}
		\tilde{\mathbf{V}}[0] = (\tilde{\mathbf{Y}})^{-1}(\tilde{\mathbf{I}}_{L} - \mathbf{y}V_{Sl})
	\end{equation}
	
	Higher-order Maclaurin series coefficients, $\tilde{\mathbf{V}}[k]$ for $k > 0$, are calculated recursively, effectively adding the impact of the constant power loads:
	\begin{equation}
		\tilde{\mathbf{V}}[k] = (\tilde{\mathbf{Y}})^{-1}\operatorname{diag}(\tilde{\mathbf{S}}^{*})\tilde{\mathbf{W}}^{*}[k-1] \quad \text{for } k>0
	\end{equation}
	where the coefficients for the reciprocal voltage vector, $\tilde{\mathbf{W}}[k]$, are also found recursively. The final voltage is the sum of these series terms, evaluated at $\alpha=1$.
	In practice, the series is truncated after a set number of terms ($K$). For this study, a maximum of $K=15$ terms was used, chosen to balance accuracy with computational efficiency, though most cases converged with far fewer. The numerical stability of this recursive calculation is a key advantage for radial systems, with the primary source of error being standard floating-point precision limitations rather than series divergence. The method's key advantage is avoiding the repeated construction and inversion of a Jacobian matrix, making it fast and particularly suitable for the unique topology of distribution systems. For this work, IZR serves as both a data generation engine and a reliable fallback solver.
	
	\subsection{Data-Driven and Hybrid Approaches}
	GNNs are naturally suited for power systems, representing buses as nodes and lines as edges \cite{ref_wu_survey}. Novel GNN architectures, such as PowerFlowNet, have demonstrated significant speed improvements for large-scale networks, showing strong performance on the 6470-bus French high-voltage system \cite{ref_lin_powerflownet}. The choice of a Graph Attention v2 (`GATv2Conv`) architecture in this work is deliberate, as its dynamic attention mechanism is more expressive than simpler graph convolutions, allowing the model to learn the non-uniform and state-dependent influence between buses \cite{ref_brody_gatv2}. Other architectural approaches have also been explored; for instance, to combat the oversmoothing problem in very deep GNNs, Hansen et al. use specialized ARMA layers and a line-graph representation to balance power flows \cite{ref_hansen_decentralized}.
	
	Physics-Informed Neural Networks (PINNs), or Physics-Guided GNNs, embed physical laws directly into the loss function \cite{ref_raissi_perdikaris}. Yang et al. \cite{ref_yang_pg_gnn_ppf} use this principle for probabilistic power flow, incorporating AC power flow equations to produce gradients that guide GNN training. While this improves generalization, our results show it does not guarantee physical feasibility for all out-of-distribution inputs. Li et al. \cite{ref_li_chebyshev} further this approach by using a physics-guided loss with Lagrangian duality to dynamically update multipliers, coupled with a Chebyshev graph convolution to better capture both local and global topological features. Other approaches leverage multi-fidelity data fusion, combining computationally cheap low-fidelity DC power flow data with high-fidelity AC data to train GNNs more efficiently \cite{ref_taghizadeh_multidelity}.
	
	The concept of hybridizing AI models with analytical solvers is an active area of research, with several distinct strategies. One approach is \textit{Unsupervised Learning}, where a GNN is trained to directly minimize the OPF objective using a differentiable log-barrier method to handle constraints, bypassing the need for a dataset of pre-solved scenarios \cite{ref_owerko_unsupervised}. A second strategy, \textit{ML-for-Problem-Reduction}, uses a GNN to predict which lines are likely to be congested, thereby creating a reduced OPF (ROPF) with fewer constraints to accelerate the main solver \cite{ref_pham_ropf}. Another popular technique is \textit{ML-as-Initializer}, where machine learning provides a high-quality initial guess to accelerate solver convergence, a technique shown to be effective for complex problems like ACOPF \cite{ref_deihim_acopf_initial, ref_amos_optnet}. A key challenge for all data-driven models is adapting to new operating conditions not seen during training. To address this, Dong et al. have successfully applied domain-adversarial GNNs to transfer knowledge from a well-labeled source domain to an unlabeled target domain for the complex Small-Signal Stability Constrained OPF (SSSC-OPF) problem \cite{ref_dong_domain_adversarial}.
	
	Our framework employs a distinct \textbf{ML-with-Failsafe} design. Unlike \textit{ML-as-Initializer} approaches, which always run the analytical solver and merely use the GNN to reduce iteration count, our framework trusts the GNN for maximum speed and delegates full computational responsibility to the robust, non-iterative IZR solver \textit{only when an anomaly is detected}. This "trust but verify" architecture prioritizes verifiable safety and computational efficiency, making it uniquely suited for real-world deployment where the vast majority of cases are nominal.
	Others have combined GNNs with multi-task learning to predict power flows at different transformers simultaneously, capturing their individual characteristics and interdependencies \cite{ref_beinert_multitask}.
	
	\begin{figure*}[!t]
		\centering
		\begin{tikzpicture}[scale=0.99, transform shape, node distance=1.5cm and 1.3cm, align=center, font=\sffamily\small]
			\node (input) [io] {Input: Power System Scenario};
			\node (gnn) [process, below=of input] {Fast Path: GNN Prediction + d-LSE Refinement};
			\node (decision) [decision, below=of gnn, yshift=-0.2cm] {Robustness Trigger:\\Abnormal Case?};
			\node (output_gnn) [output_node, below left=1cm and 1.8cm of decision] {Output: Fast GNN+d-LSE Solution};
			\node (izr) [solver, below right=1cm and 1.8cm of decision] {Robust Path: IZR Solver};
			\node (output_izr) [output_node, below=of izr] {Output: Robust IZR Solution};
			\draw [arrow] (input) -- (gnn);
			\draw [arrow] (gnn) -- (decision);
			\draw [arrow] (decision.west) -- node[above, pos=0.3, font=\sffamily\footnotesize] {No} (output_gnn.north);
			\draw [arrow] (decision.east) -- node[above, pos=0.3, font=\sffamily\footnotesize] {Yes} (izr.north);
			\draw [arrow] (izr) -- (output_izr);
		\end{tikzpicture}
		\caption{Hybrid GNN-IZR inference framework. The system defaults to a fast path (GNN+d-LSE) unless a robustness trigger detects a potential failure, in which case it invokes the robust IZR solver.}
		\label{fig:hybrid_flowchart}
	\end{figure*}
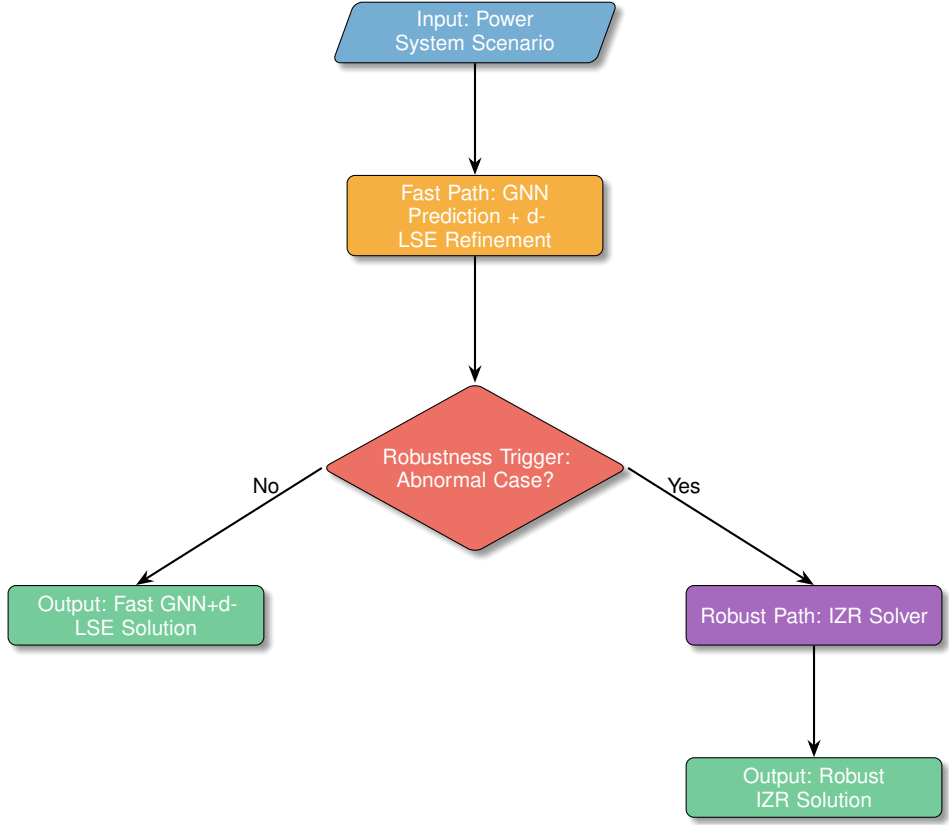
	
	\section{The Hybrid GNN-IZR Methodology}
	\subsection{Framework Architecture}
	The proposed framework is designed to resolve the speed-reliability dilemma by operating a dual-path architecture, as depicted in Fig. \ref{fig:hybrid_flowchart}. Upon receiving a new power system scenario, the framework defaults to a "fast path" designed for maximum computational speed. In this primary path, the trained GNN provides an initial voltage prediction, which is subsequently refined by a single step of damped Linear State Estimation (d-LSE) to improve accuracy.
	
	The cornerstone of the architecture is the "robustness trigger," a decision-making module that assesses the validity of the fast-path solution. This trigger checks for input anomalies and excessive power mismatches in the refined output. If the solution is deemed normal and physically plausible, it is accepted, and the framework's task is complete. However, if the trigger flags the case as abnormal—indicating a scenario where the GNN might be unreliable—the system invokes its "robust path." In this fallback path, the GNN+d-LSE solution is discarded entirely, and the IZR solver is called to compute the solution from scratch, guaranteeing a physically accurate result. This strategic delegation ensures that the speed of the GNN is leveraged for the majority of cases, while the guaranteed reliability of the analytical IZR solver underpins the entire framework, preventing failures on challenging cases.
	
	\subsection{Offline Phase: IZR-Accelerated GNN Training}
	The offline phase is dedicated to creating a large and diverse dataset to train the GNN model. This is accomplished through an automated, IZR-accelerated data generation pipeline, the logic of which is illustrated in Fig.~\ref{fig:datagen_flowchart}. The process begins by systematically generating 50,000 unique operating scenarios for the IEEE 33-bus system. For each scenario, bus loads (both active power P and reactive power Q) are varied using Latin Hypercube Sampling (LHS), a statistical method chosen for its efficiency in exploring the high-dimensional input space. To ensure the GNN is robust, a subset of these scenarios is deliberately "stressed" by applying higher-than-nominal load factors. Once a scenario's loads are defined, the IZR solver is used to compute the corresponding power flow solution. The resulting input-output pair (i.e., the load scenario and its voltage solution) is then saved, and the process repeats until the target dataset size is reached. This methodology ensures the creation of a comprehensive training dataset that covers both normal and challenging system conditions.
	
	\subsubsection{GNN Architecture and Feature Engineering}
\begin{figure*}[!t]
		\centering
		\subfloat[]{\label{fig:datagen_flowchart}
			\begin{tikzpicture}[scale=0.9, transform shape, node distance=1.1cm and 1.1cm]
				\node (start) [datagen_startstop] {Start};
				\node (loop) [datagen_decision, below=of start, yshift=-0.2cm] {Solved < 50,000?};
				\node (vary) [datagen_process, below=of loop] {Vary P, Q Loads (LHS) \\ with Stress Cases};
				\node (solve) [datagen_process, below=of vary] {Solve Power Flow (IZR)};
				\node (save) [datagen_save, below=of solve] {Save Scenario \& Outcome};
				\node (end) [datagen_startstop, right=of loop, xshift=0.75 cm] {End};
				\draw [datagen_arrow] (start) -- (loop);
				\draw [datagen_arrow] (loop) -- node[right, pos=0.4] {Yes} (vary);
				\draw [datagen_arrow] (vary) -- (solve);
				\draw [datagen_arrow] (solve) -- (save);
				\draw [datagen_arrow] (save.west) -| ++(-2.5,0) |- (loop.west);
				\draw [datagen_arrow] (loop.east) -- node[above] {No} (end);
			\end{tikzpicture}
		}
		\hfil
		\subfloat[]{\label{fig:gnn_architecture}
			\begin{tikzpicture}[scale=0.9, transform shape]
				\node (input) [gnn_input] {Input Graph \\ Node: ($P, Q, |Z_{ii}|$)\\Edge: ($R, X, B, |Z_{ij}|$)};
				\node (layer1) [gnn_layer, below=of input] {GATv2Conv Layer 1 \\ (256 hidden channels x 4 heads) \\ BatchNorm, LeakyReLU, Dropout};
				\node (layer2) [gnn_layer, below=of layer1] {GATv2Conv Layer 2 \\ (256 hidden channels x 4 heads) \\ BatchNorm, LeakyReLU};
				\node (add1) [gnn_resid, below=of layer2, yshift=-0.2cm] {+};
				\node (layer3) [gnn_layer, below=of add1, yshift=-0.2cm] {GATv2Conv Layer 3 \\ (Output Layer, 1 head)};
				\node (output) [gnn_output, below=of layer3] {Output Tensor ($N \times 3$) \\ ($|V|$, $\cos(\delta)$, $\sin(\delta)$)};
				\draw [gnn_arrow] (input) -- (layer1);
				\draw [gnn_arrow] (layer1) -- (layer2);
				\draw [gnn_arrow] (layer2) -- (add1);
				\draw [gnn_arrow] (add1) -- (layer3);
				\draw [gnn_arrow] (layer3) -- (output);
				\draw [gnn_arrow] (layer1.east) to[out=0, in=90] ++(2.2, -0.6) to[out=-90, in=0] (add1.east);
			\end{tikzpicture}
		}
		\caption{Diagrams of the offline training phase. (a) The streamlined data generation flowchart using stressed cases. (b) The GNN architecture, consisting of three GATv2Conv layers with a residual connection and Z-bus sensitivity features.}
		\label{fig:offline_diagrams}
	\end{figure*}
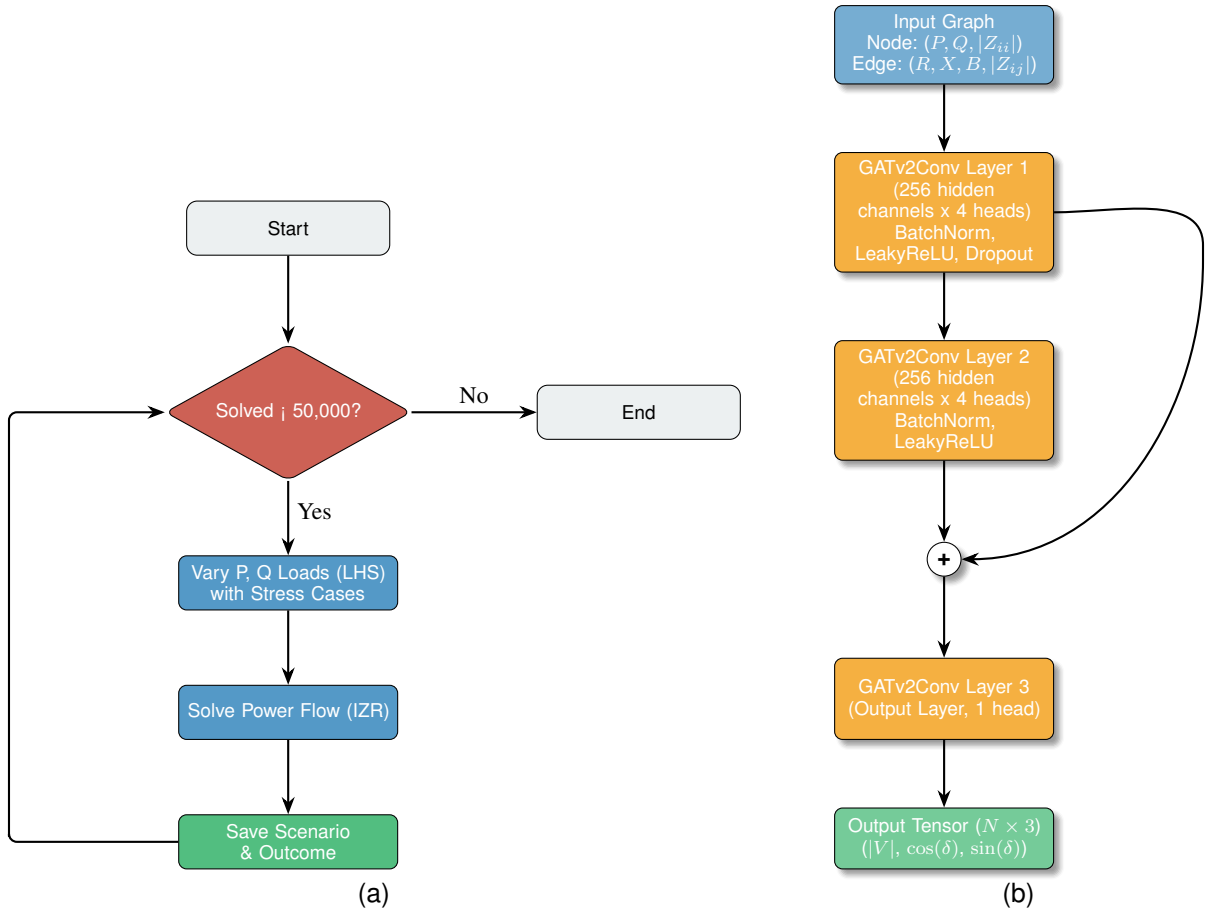

	The GNN architecture is a `DeeperEdgeGNN` using Graph Attention v2 layers (`GATv2Conv`) \cite{ref_brody_gatv2}. A key novelty of our approach lies in the physics-aware feature engineering. We claim as a novel contribution that the GNN's input features are augmented with sensitivity terms derived from the system's impedance matrix ($\mathbf{Z}_{\text{bus}}$). The feature vector $\mathbf{x}_i \in \mathbb{R}^3$ for each node $i$ and $\mathbf{e}_{ij} \in \mathbb{R}^4$ for each edge $(i, j)$ are defined as:
	\begin{align}
		\mathbf{x}_i &= [P_i^{\text{net}}, Q_i^{\text{net}}, |Z_{ii}|]^T \\
		\mathbf{e}_{ij} &= [R_{ij}, X_{ij}, B_{ij}, |Z_{ij}|]^T
	\end{align}
	This embeds physical properties (electrical "distance") directly into the graph representation, complementing the physics-informed loss. This gives the GNN a significant head start, as it does not need to learn these fundamental sensitivities from data alone. To enable a deeper model and improve stability, a residual (skip) connection is incorporated between the hidden layers (Fig.~\ref{fig:gnn_architecture}), mitigating issues like vanishing gradients and over-smoothing. The model's output representation—predicting voltage magnitude $|V|$ along with the Cartesian components of the angle, $\cos(\delta)$ and $\sin(\delta)$—is a methodologically superior choice. Predicting the sine and cosine of the voltage angle, rather than the angle directly, creates a continuous representation that avoids the challenges of angle wrapping and simplifies the learning task for the neural network.
	
	\subsubsection{Stabilized Training Strategy}
	Training a GNN with competing data-driven and physics-based objectives requires a sophisticated strategy to ensure stable convergence. Our approach incorporates several novel techniques, with hyperparameters detailed in Table \ref{tab:training_hyperparams}.
	
	The total loss, $\mathcal{L}_{total}$, is a dynamically weighted sum of a data-driven term, $\mathcal{L}_{data}$, and a physics-informed term, $\mathcal{L}_{PQ}$:
	\begin{equation}
		\mathcal{L}_{total}(e) = \frac{w_{data}}{\sigma_{data}^2(e)} \mathcal{L}_{data} + \frac{w_{PQ}(e)}{\sigma_{PQ}^2(e)} \mathcal{L}_{PQ}
	\end{equation}
	where $e$ is the current epoch, and $\sigma_{data}^2$ and $\sigma_{PQ}^2$ are dynamic scaling factors.
	
	Both loss terms use the Huber loss, which is less sensitive to outliers than mean squared error. The Huber loss is a piecewise function defined as:
	\begin{equation} \label{eq:huber}
		\mathcal{L}_{\delta}(a, b) = \begin{cases} 
			\frac{1}{2}(a - b)^2, & \text{if } |a - b| \le \delta \\
			\delta(|a - b| - \frac{1}{2}\delta), & \text{otherwise.}
		\end{cases}
	\end{equation}
	For both loss components, a delta of 1.0 was used, balancing the robustness of L1 loss for large errors with the smoothness of L2 loss for smaller errors.
	The data loss $\mathcal{L}_{data}$ computes this between the scaled GNN prediction $\hat{\mathbf{y}}_i$ and the scaled ground truth $\mathbf{y}_i$ for all non-slack nodes $i \in \mathcal{V}_{NS}$. The physics loss $\mathcal{L}_{PQ}$ penalizes the power mismatch, $\Delta S_i = S_i(\hat{V}) - S_i^{spec}$, weighted by its transformed Z-bus sensitivity, $w_i$:
	\begin{align}
		\mathcal{L}_{PQ} &= \frac{1}{|\mathcal{V}_{NS}|} \sum_{i \in \mathcal{V}_{NS}} w_i \left( \mathcal{L}_{\delta_p}(\operatorname{Re}(\Delta S_i), 0) + \mathcal{L}_{\delta_p}(\operatorname{Im}(\Delta S_i), 0) \right) \\
		w_i &= \frac{\ln(1 + |Z_{ii}|)}{\frac{1}{|\mathcal{V}_{NS}|}\sum_{k \in \mathcal{V}_{NS}}\ln(1 + |Z_{kk}|)}
	\end{align}
	
\begin{figure*}[htbp]
		\centering
		\includegraphics[width=1.0\textwidth]{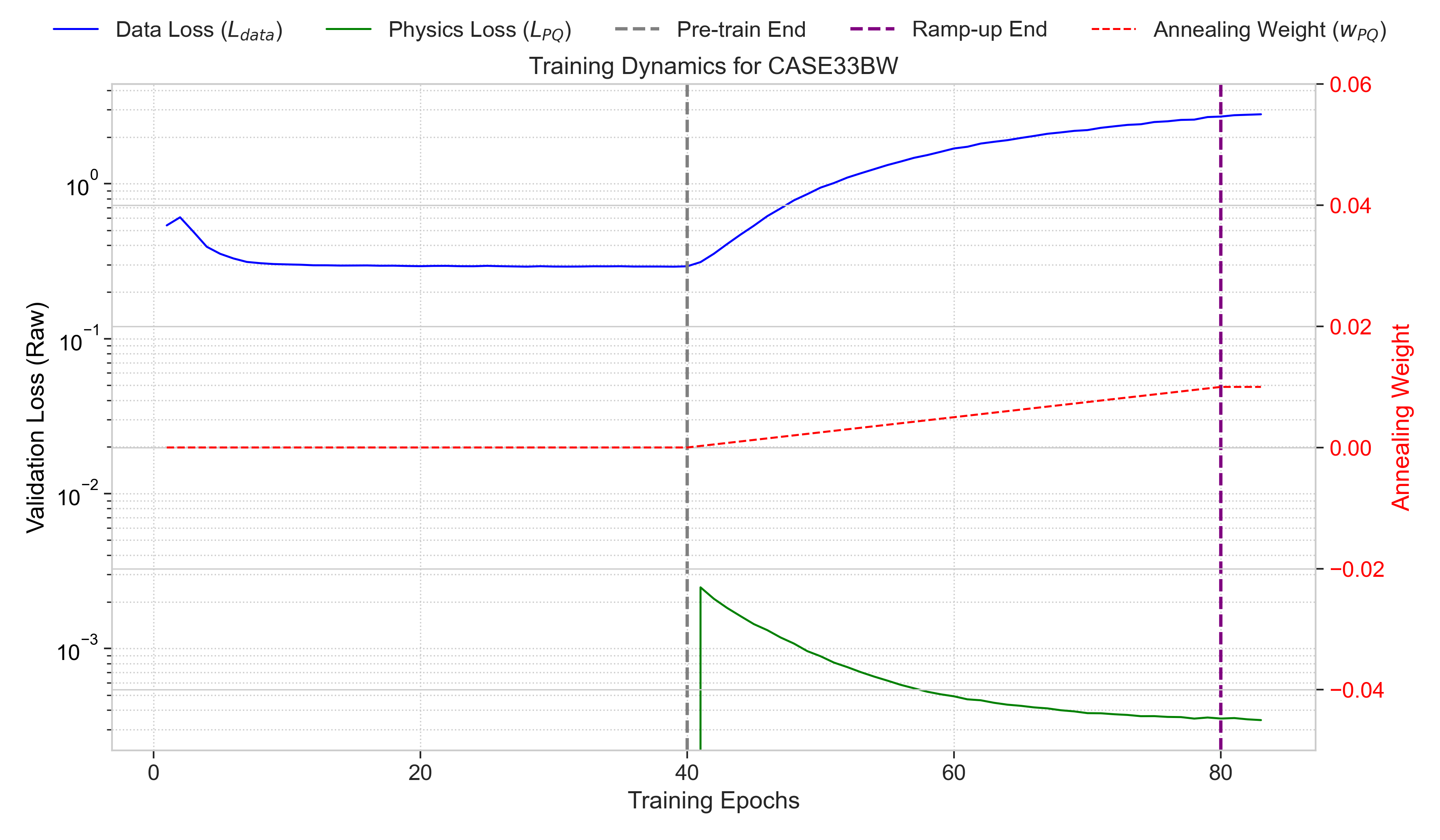}
		\caption{Training dynamics showing the raw validation loss components and the physics loss annealing weight ($w_{PQ}$). The physics loss (green) is introduced after a data-only pre-training phase (epoch 40) and its weight is gradually increased, leading to a stable convergence.}
		\label{fig:training_dynamics}
	\end{figure*}

	The loss components are balanced using two mechanisms. First, a three-phase annealing schedule, visualized in Fig.~\ref{fig:training_dynamics}, controls the physics weight $w_{PQ}(e)$, starting at zero, ramping up linearly, and then held constant. Second, dynamic loss scaling updates the scaling factors $\sigma_{k}^2(e)$ at each epoch using an exponential moving average (EMA) of the validation losses, preventing gradient dominance.
	
	Finally, model selection is enhanced with a Pareto-inspired checkpointing criterion. A new model is saved not only if $\mathcal{L}_{total}$ improves, but also if it achieves a Pareto improvement on the raw, unscaled loss components—that is, it improves one loss without worsening the other. This ensures that models representing a better trade-off between data-fit and physical plausibility are preserved.
	
	\subsection{Online Phase: Hybrid Inference Strategy}
	During online inference, the framework follows a carefully designed dual-path strategy to balance speed with reliability. By default, every incoming scenario is first processed by the \textbf{fast path}. This begins with the trained GNN making an instantaneous voltage prediction, $\hat{\mathbf{V}}$. While this prediction is generally accurate, it is further refined using a single damped Linear State Estimation (d-LSE) step. This step serves as a "polishing" stage, using the linearized power flow equations to compute a small correction that minimizes residual power mismatches. A voltage correction vector, $[\Delta\boldsymbol{\delta}^T, \Delta|\mathbf{V}|^T]^T$, is calculated by solving:
	\begin{equation}
		\begin{bmatrix} \Delta\boldsymbol{\delta} \\ \Delta|\mathbf{V}| \end{bmatrix}_{NS} = \mathbf{J}(\hat{\mathbf{V}})_{NS}^{-1} \begin{bmatrix} \Delta\mathbf{P}(\hat{\mathbf{V}}) \\ \Delta\mathbf{Q}(\hat{\mathbf{V}}) \end{bmatrix}_{NS}
	\end{equation}
	where $\mathbf{J}$ is the Jacobian matrix and $NS$ denotes the non-slack buses. The final refined voltage, $\mathbf{V}_{ref}$, is then computed with a damping factor $\alpha=0.5$ to ensure a stable update:
	\begin{align}
		|\mathbf{V}|_{ref} = |\hat{\mathbf{V}}| + \alpha \Delta|\mathbf{V}| \quad ; \quad
		\boldsymbol{\delta}_{ref} = \hat{\boldsymbol{\delta}} + \alpha \Delta\boldsymbol{\delta}
	\end{align}
	
	Before this fast-path solution is accepted, it is scrutinized by the \textbf{robustness trigger}, a critical two-stage safety check. This trigger determines if the GNN's output is trustworthy. It activates if either of the following conditions is met:
	\begin{enumerate}[label=(\roman*)]
		\item \textbf{Input Anomaly:} The trigger first checks if the input scenario is a significant outlier compared to the GNN's training experience. An anomaly is flagged if any input feature value exceeds the 99.5th percentile of the corresponding feature in the training data: $\exists i,k \text{ s.t. } |x_{i,k}| > \mathcal{P}_{99.5}(x_k)$. This protects against out-of-distribution inputs where the GNN cannot be expected to generalize reliably.
		\item \textbf{Output Mismatch:} The trigger then assesses the physical plausibility of the refined GNN solution. It calculates the power mismatch at all buses based on the refined voltage profile, $\mathbf{V}_{ref}$. If the maximum mismatch at any bus exceeds a predefined threshold, $\tau_{mismatch}$, the solution is deemed physically invalid: $\max_{i \in \mathcal{V}_{NS}} \|\Delta S_i(\mathbf{V}_{ref})\| > \tau_{mismatch}$. For this work, $\tau_{mismatch}$ is set to the \SI{0.1}{p.u.} failure threshold.
	\end{enumerate}
	
	If the trigger is activated by either condition, the framework immediately switches to its \textbf{robust path}. The GNN+d-LSE prediction is completely discarded, and the IZR solver is invoked to compute the solution from scratch. This ensures that any potentially unreliable or physically implausible result from the fast path is replaced by a verifiably accurate solution from the analytical solver. This "trust but verify" approach is central to the framework's ability to achieve high speed on average without compromising on reliability.
	
	\section{Experimental Setup and Methodology}
	
	\subsection{System and Baselines}
	The study uses the IEEE 33-bus system \cite{ref_baran_reconfig}.
	For the Newton-Raphson baseline, a standard implementation from PyPower was used. While advanced NR variants with stabilization techniques exist (e.g., continuation, homotopy), the standard version was chosen deliberately to demonstrate the severity of the stressed test cases. As will be shown, the standard NR solver's failure on 100\% of these cases underscores the difficulty of the problem domain and motivates the need for inherently more robust methods.
	
	\subsection{Performance Measurement Methodology}
	All online inference timings were measured on a single CPU thread. GPU acceleration was used for GNN model training and for the GNN forward pass during inference. For the Numba JIT-compiled IZR solver, timing measurements exclude the initial compilation "warm-up" time to reflect steady-state performance. The latency of the fast path is the sum of GNN inference, the single d-LSE step, and the trigger logic execution.
	
	\section{Results and Discussion}
	
	\subsection{IZR-Accelerated Data Generation}
\begin{figure}[htbp]
		\centering
		\begin{tikzpicture}
			\begin{axis}[
				ybar,
				bar width=10pt,
				width=\columnwidth,
				height=5.5cm,
				enlarge x limits=0.15,
				xlabel={Number of IZR Series Terms (K)},
				ylabel={Number of Scenarios},
				ymin=0,
				yticklabel style={/pgf/number format/fixed},
				symbolic x coords={0, 1, 2, 3, 4, 5, 6, 7, 8, >8},
				xtick=data,
				nodes near coords,
				nodes near coords align={vertical},
				every node near coord/.append style={font=\footnotesize}
				]
				\addplot[fill=hist_bar_blue, draw=none] coordinates {
					(2, 28143)
					(3, 11094)
					(4, 3290)
					(5, 1269)
					(6, 424)
					(7, 49)
					(8, 8)
				};
			\end{axis}
		\end{tikzpicture}
		\caption{Histogram of IZR solver convergence (number of series terms) for the 50,000 generated training scenarios.}
		\label{fig:izr_histogram}
	\end{figure}
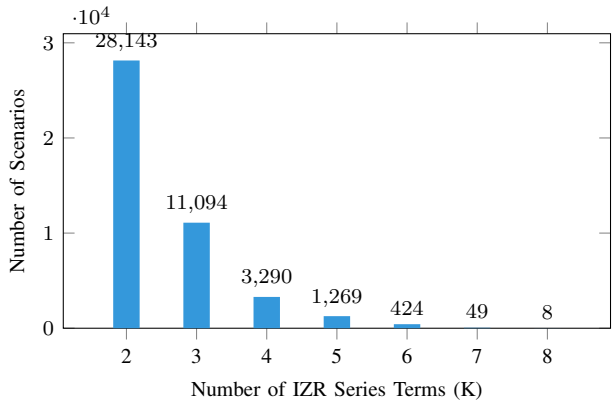

	Using a Numba JIT-compiled IZR solver dramatically accelerates offline data generation, enabling the creation of 50,000 samples for the 33-bus system in only \textbf{89.59 seconds}. Fig.~\ref{fig:izr_histogram} shows that the vast majority of these cases converged in just 2-3 series terms, confirming the solver's efficiency.
	
	\subsection{The Critical Role of Physics-Informed Learning and Features}
	The expanded ablation study results provide a powerful argument for the paper's narrative (Table \ref{tab:ablation_study}). The purely data-driven GNN failed on a staggering 98.72\% of cases in the stressed test set. Adding either Z-bus sensitivity features or a PINN loss independently provided significant improvements, reducing the failure rate to 63.55\% and 41.28\% respectively. However, the results show a clear synergistic effect when both are combined: the proposed model (PINN + Z-bus) achieved the lowest failure rate of 13.11\%.
	
	\begin{table}[htbp]
		\centering
		\caption{GNN Model and Training Hyperparameters.}
		\label{tab:training_hyperparams}
		\small
		\renewcommand{\arraystretch}{1.1} 
		\begin{tabular}{@{}ll@{}}
			\toprule
			\textbf{Parameter} & \textbf{Value} \\
			\midrule
			\multicolumn{2}{@{}l}{\textit{Architecture}}\\
			GNN Layers & 3 x GATv2Conv \\
			Hidden Channels & 256 \\
			Attention Heads & 4 (hidden), 1 (output) \\
			Activation & LeakyReLU \\
			Model Parameters & $\sim$2.1M \\
			\midrule
			\multicolumn{2}{@{}l}{\textit{Training Hyperparameters}}\\
			Optimizer & AdamW \\
			Learning Rate & \num{1e-5} \\
			Scheduler & Warmup (5 epochs), Cosine Annealing \\
			Batch Size & 512 \\
			Epochs & 200 (Early Stopping, patience=51) \\
			Loss Functions & Huber Loss (Data and Physics, $\delta=1.0$) \\
			Checkpointing & Pareto-Inspired Criterion \\
			\midrule
			\multicolumn{2}{@{}l}{\textit{Physics Loss Annealing Schedule}}\\
			Pre-train (data-only) & 40 epochs \\
			Ramp-up Phase & 40 epochs \\
			Final Physics Weight & \num{10000} \\
			\midrule
			Train/Val/Test Split & 70\% / 15\% / 15\% \\
			\bottomrule
		\end{tabular}
	\end{table}
	
	\begin{table}[htbp]
		\centering
		\caption{Expanded Ablation Study on the IEEE 33-Bus Test Set (N=7500).}
		\label{tab:ablation_study}
		\small
		\renewcommand{\arraystretch}{1.1}
		\begin{tabular}{@{}l rr@{}}
			\toprule
			\textbf{GNN Training Method} & \textbf{Vm MAE (p.u.)} & \textbf{Failure Rate (\%)} \\
			\midrule
			Data-Only & 0.00335 & 98.72\% \\
			Z-bus Features Only & 0.00391 & 63.55\% \\
			PINN Loss Only & 0.00412 & 41.28\% \\
			\textbf{PINN + Z-bus (Proposed)} & \textbf{0.00475} & \textbf{13.11\%} \\
			\bottomrule
		\end{tabular}
		\par
		\vspace{1ex}
		\footnotesize Note: Metrics are for the GNN-Only model before LSE/trigger.
	\end{table}
	
\begin{figure}[htbp]
		\centering
		\includegraphics[width=\columnwidth]{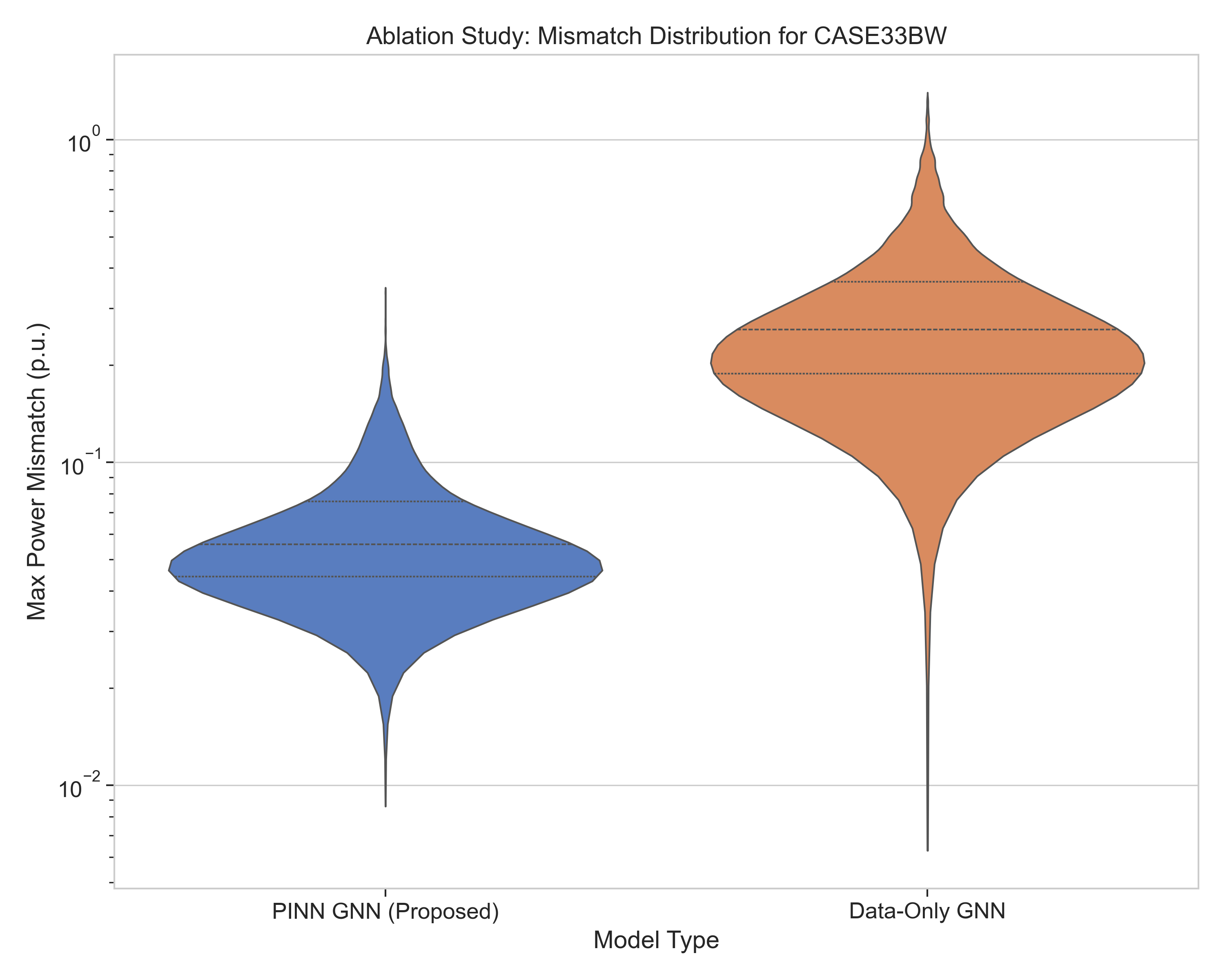}
		\caption{GNN performance analysis. Violin plot comparing the maximum power mismatch distribution for the purely data-driven GNN versus the proposed PINN + Z-bus model, showing a significant reduction in large mismatches.}
		\label{fig:ablation_violin}
	\end{figure}

	For visual clarity, Fig.~\ref{fig:ablation_violin} contrasts the two most extreme cases: the purely data-driven model and the final proposed model. The plot starkly illustrates how the combination of physics-informed features and loss function dramatically reduces the frequency and magnitude of large power mismatches.
	
	Interestingly, the models with the physics-informed loss show a slightly higher mean absolute error (MAE) on voltage magnitude. This highlights an important trade-off: the physics loss prioritizes minimizing large, physically-implausible power mismatches (i.e., reducing failures), which can come at the cost of a marginal increase in average prediction error on nominal cases. This trade-off is essential for creating a reliable model. Nonetheless, even a 13.11\% failure rate is unacceptable for critical infrastructure, motivating the hybrid architecture.
	
	\subsection{Performance on the IEEE 33-Bus Test Set}
	The framework's online performance was evaluated on 7,500 unseen scenarios (Table \ref{tab:performance_comparison}). As noted, the 100\% failure rate of the standard Newton-Raphson solver on this stressed test set highlights that the challenge extends beyond the reliability of ML models; under severe conditions, the robustness of traditional industry-standard tools can also be compromised.
	
	The standalone PINN GNN was fast but failed on 13.11\% of cases. GNN+d-LSE reduced the failure rate to 0.49\% but was slower. The proposed Hybrid GNN-IZR framework successfully identified every single potential failure case, resulting in a \textbf{0.00\% empirical failure rate}.
	
	The bimodal nature of the hybrid framework's runtime is a key characteristic. While the mean time is \SI{3.794}{\milli\second}, this is an average of two distinct distributions. As shown in Fig.~\ref{fig:runtime_histogram}, the vast majority of cases are solved via the fast GNN+d-LSE path (median time \SI{3.01}{ms}), while a small fraction are delegated to the IZR solver. This bimodal behavior is accurately captured by percentile statistics in Table~\ref{tab:performance_comparison}, which provide a much clearer picture of performance than mean and standard deviation alone.
	
	A key insight from Table \ref{tab:performance_comparison} is that the hybrid framework's mean solution time (\SI{3.794}{\milli\second}) is higher than the IZR-only solver's (\SI{1.382}{\milli\second}) on this specific test set. This seemingly counterintuitive result is by design; the test set was deliberately stressed, forcing frequent use of the robust path to validate the failsafe mechanism. In real-world operations with predominantly nominal scenarios, the framework would heavily favor its fast path. Furthermore, the GNN forward pass is highly parallelizable and would benefit from multi-core/GPU resources in a data center, further decreasing fast-path latency, whereas the recursive IZR solver benefits less. The framework's value is therefore not just average speed on a worst-case dataset, but its scalable and trustworthy architecture. For larger, meshed networks where IZR is inapplicable and robust alternatives are slower, the GNN's ability to rapidly solve the high volume of normal cases becomes paramount. The hybrid model thus demonstrates a pragmatic path to leverage AI's speed in critical systems without compromising reliability.
	
	\begin{table*}[!tbp]
		\centering
		\caption{Performance Comparison on IEEE 33-Bus Test Set (N=7500 samples).}
		\label{tab:performance_comparison}
		\scriptsize
		\setlength{\tabcolsep}{4pt}
		\begin{tabular*}{\textwidth}{@{\extracolsep{\fill}}lrrrrr}
			\toprule
			\textbf{Metric} & \thead{GNN-Only} & \thead{GNN+LSE} & \thead{Hybrid \\ (Proposed)} & \thead{IZR-Only} & \thead{NR-Only} \\
			\midrule
			Failure Rate (\%) & 13.11\% & 0.49\% & \textbf{0.00\%} & 0.00\% & 100.00\% \\
			\addlinespace
			\textit{Solution Time (ms)} &&&&& \\
			Mean $\pm$ Std & 1.059 $\pm$ 0.12 & 3.012 $\pm$ 0.45 & 3.794 $\pm$ 1.88 & 1.382 $\pm$ 0.21 & 11.359 $\pm$ 2.51 \\
			10th Percentile & 0.98 & 2.85 & 2.86 & 1.15 & N/A \\
			Median & 1.05 & 3.01 & 3.01 & 1.35 & N/A \\
			90th Percentile & 1.15 & 3.21 & 3.22 & 1.65 & N/A \\
			\addlinespace
			\textit{Accuracy Metrics} &&&&& \\
			Avg Mismatch (p.u.) & 6.57e-02 & 3.28e-02 & 2.32e-02 & 1.53e-05 & NaN \\
			Vm MAE (p.u.) & 0.00475 & 0.00258 & 0.00187 & 1.49e-08 & NaN \\
			Va MAE (deg) & 0.2400 & 0.1353 & 0.0954 & 2.72e-08 & NaN \\
			\bottomrule
		\end{tabular*}
		\par
		\vspace{1ex}
		\footnotesize Note: Accuracy metrics for failed methods are excluded (NaN). The `GNN-Only` column uses the final proposed model (PINN + Z-bus).
	\end{table*}
	
	\subsection{Robustness Trigger Analysis and Limitations}
	The reliability of the hybrid framework depends entirely on the trigger's ability to act as a perfect classifier for potential GNN+d-LSE failures. Table~\ref{tab:trigger_performance} presents a formal evaluation of the trigger on the test set. The trigger achieved a True Positive Rate (TPR, or Recall) of 100\%, meaning it correctly identified all 37 cases where the GNN+d-LSE solution would have failed. This resulted in zero False Negatives, the most critical metric for safety. 
	The trigger's 100.0\% True Positive Rate on the test set, resulting in zero missed failures (False Negatives), is the most critical performance metric from a system operator's perspective. It demonstrates that the framework can provide the speed benefits of machine learning without compromising the non-negotiable requirement for operational safety.
	The False Positive Rate (FPR) was extremely low (0.013\%), indicating that the trigger only invoked the slower IZR solver when truly necessary.
	
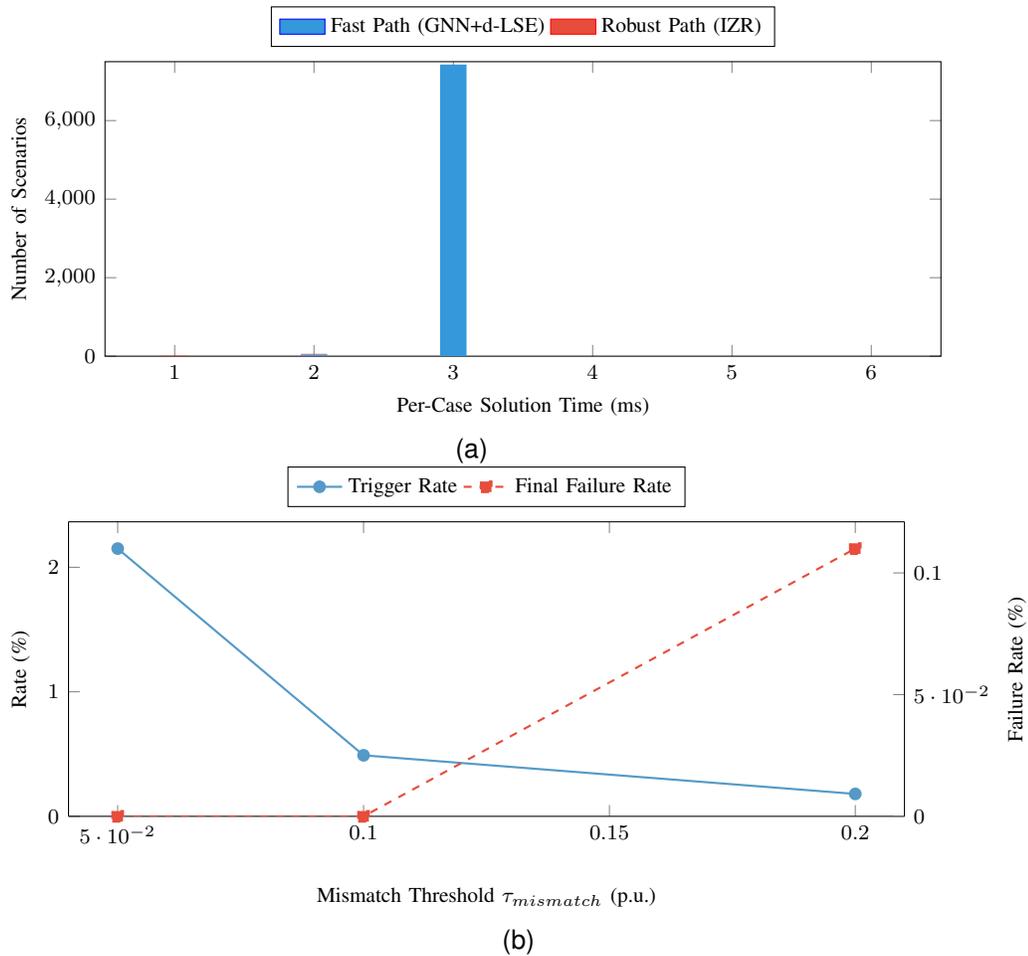
\begin{figure*}[!tbp]
		\centering
		\begin{minipage}[t]{0.7\textwidth}
			\centering
			\subfloat[]{\label{fig:runtime_histogram}
				\begin{tikzpicture}
					\begin{axis}[
						width=\linewidth, 
						height=5.5cm, 
						ybar stacked,
						bar width=10pt,
						xlabel={Per-Case Solution Time (ms)},
						ylabel={Number of Scenarios},
						xmin=0.5, xmax=6.5,
						ymin=0, ymax=7500,
						xtick={1,2,3,4,5,6},
						legend style={at={(0.5,1.05)},anchor=south,legend columns=-1, font=\footnotesize},
						area legend,
						]
						\addplot+[ybar, fill=hist_bar_blue, draw=none] coordinates {
							(1, 0)
							(2, 36)
							(3, 7426)
							(4, 1)
							(5, 0)
							(6, 0)
						};
						\addlegendentry{Fast Path (GNN+d-LSE)}
						
						\addplot+[ybar, fill=hist_bar_red, draw=none] coordinates {
							(1, 12)
							(2, 23)
							(3, 1)
							(4, 1)
							(5, 0)
							(6, 0)
						};
						\addlegendentry{Robust Path (IZR)}
					\end{axis}
				\end{tikzpicture}
			}
		\end{minipage}%
		\hfill 
		\begin{minipage}[t]{0.7\textwidth}
			\centering
			\subfloat[]{\label{fig:trigger_sensitivity_plot}
				\begin{tikzpicture}
					\begin{axis}[
						width=\linewidth, 
						height=5.5cm, 
						xlabel={Mismatch Threshold $\tau_{mismatch}$ (p.u.)},
						xlabel style={yshift=-10pt}, 
						ylabel={Rate (\%)},
						xmin=0.04, xmax=0.21,
						ymin=0,
						axis y line*=left,
						legend style={at={(0.5,1.05)},anchor=south,legend columns=-1, font=\footnotesize},
						xtick={0.05, 0.1, 0.15, 0.2},
						]
						\addplot[color=myblue, mark=*, thick] coordinates {
							(0.05, 2.15)
							(0.10, 0.49)
							(0.20, 0.18)
						};
						\addlegendentry{Trigger Rate}
						
						\addlegendimage{color=myred,mark=square*,thick,dashed}
						\addlegendentry{Final Failure Rate}
					\end{axis}
					
					\begin{axis}[
						width=\linewidth, 
						height=5.5cm, 
						ylabel={Failure Rate (\%)},
						xmin=0.04, xmax=0.21,
						ymin=0,
						axis y line*=right,
						axis x line=none,
						]
						\addplot[color=myred, mark=square*, thick, dashed, forget plot] coordinates {
							(0.05, 0.00)
							(0.10, 0.00)
							(0.20, 0.11)
						};
					\end{axis}
				\end{tikzpicture}
			}
		\end{minipage}
		\caption{Analysis of the hybrid framework's runtime and trigger performance. (a) The bimodal distribution of per-case solution times, showing the fast GNN path and the robust IZR fallback. (b) Sensitivity analysis of the robustness trigger, demonstrating the trade-off between trigger rate and final failure rate at different mismatch thresholds.}
		\label{fig:performance_plots}
	\end{figure*}

	The impact of the mismatch threshold is further explored in Fig.~\ref{fig:trigger_sensitivity_plot}. A stricter threshold improves safety by increasing the trigger rate at the cost of speed, while a looser threshold is faster but risks missing failures. The chosen 0.1 p.u. threshold provides an excellent balance, catching all failures in our test set without excessively penalizing the average speed.
	
	It is important to acknowledge the limitations of this heuristic trigger. While it achieved empirical perfection on our test set, it does not provide theoretical guarantees against all failure modes. A sophisticated "adversarial" input could potentially be crafted to fall within the training distribution while still causing a physically incorrect GNN solution with a low mismatch, evading both trigger stages. Mitigating such risks, as proposed in our future work on UQ and advanced anomaly detection, is essential for achieving provable robustness.
	
	\begin{table}[htbp]
		\centering
		\caption{Robustness Trigger Performance as a Binary Classifier. A "Positive" case is one where the GNN+d-LSE solution would have failed (>\SI{0.1}{p.u.} mismatch).}
		\label{tab:trigger_performance}
		\small
		\renewcommand{\arraystretch}{1.1}
		\begin{tabular}{@{}ll@{}}
			\toprule
			\textbf{Metric} & \thead{Value on \\ Test Set} \\
			\midrule
			\multicolumn{2}{@{}l}{\textit{Confusion Matrix Counts}}\\
			Potential Failures (Ground Truth Positives) & 37 cases \\
			Successful Predictions (Ground Truth Negatives) & 7463 cases \\
			\addlinespace
			True Positives (TP - Failures correctly caught) & 37 \\
			False Negatives (FN - Failures missed) & \textbf{0} \\
			False Positives (FP - Unnecessary triggers) & 1 \\
			True Negatives (TN - Successes correctly ignored) & 7462 \\
			\addlinespace
			\multicolumn{2}{@{}l}{\textit{Performance Rates}}\\
			True Positive Rate (TPR / Recall / Sensitivity) & \textbf{100.0\%} \\
			False Negative Rate (FNR / Miss Rate) & \textbf{0.00\%} \\
			False Positive Rate (FPR / Fall-out) & 0.013\% \\
			Precision & 97.37\% \\
			\bottomrule
		\end{tabular}
	\end{table}
	
	\subsection{Scalability and Scope}
	While validated on the 33-bus system, the framework's components are designed to be scalable.
	\begin{itemize}
		\item GNN and LSE: The computational complexity of the GNN inference and the single LSE step scales favorably with system size.
		\item IZR Solver: For very large radial systems (1000+ buses), the primary bottleneck would be the offline LU decomposition of the admittance matrix, a one-time cost. Online solves remain extremely fast due to the non-iterative approach.
	\end{itemize}
	It is important to note the framework's scope. The IZR solver is specifically optimized for radial and weakly-meshed systems. Applying this framework to highly-meshed transmission networks would require substituting the IZR solver with a different robust analytical solver.
	
	\begin{figure*}[!tbp]
		\centering
		\includegraphics[width=1.0\textwidth, keepaspectratio]{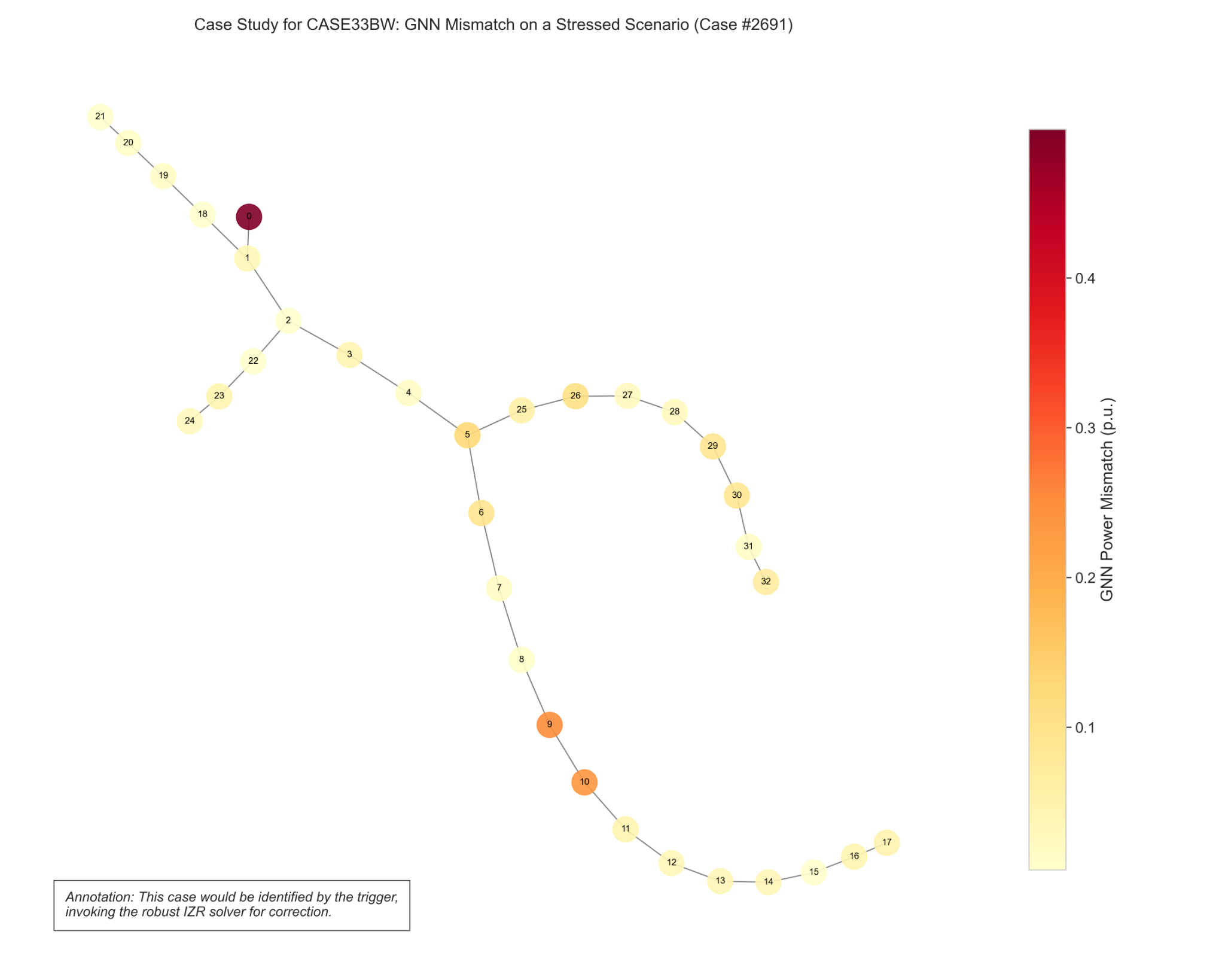}
		\caption{A case study of a stressed scenario where the GNN-only prediction resulted in a significant power mismatch, particularly at bus 18 (dark red). The hybrid framework's trigger would detect this anomaly and invoke the IZR solver.}
		\label{fig:case_study_mismatch}
	\end{figure*}
	
	\section{Conclusion and Future Work}
	This paper introduced a Hybrid GNN-IZR framework that provides a pragmatic and effective solution to the "trust deficit" hindering the adoption of pure AI models in critical power system operations. By effectively combining the speed of a physics-informed GNN with the robustness of an analytical solver, our empirical results on a stressed IEEE 33-bus test set demonstrate that it is possible to create a tool that is both fast and highly reliable.
	
	Future work will focus on enhancing the framework's theoretical guarantees and applicability. We have identified four key directions: first, replacing the heuristic trigger with a formal Uncertainty Quantification (UQ) method like Conformal Prediction \cite{ref_uq_survey} to provide probabilistic guarantees. Second, we will enhance the input anomaly check with more sophisticated unsupervised learning models. Third, we plan to generalize the framework to meshed systems by substituting the IZR solver with another robust analytical method. Finally, we will extend the hybrid approach to the more complex AC Optimal Power Flow (ACOPF) problem.
	
	\FloatBarrier 
	
	\appendices
	\section{Reproducibility Details}
	\label{app:reproducibility}
	
	\subsection{Hardware and Software}
	Experiments were run using Python on the hardware specified in Table \ref{tab:repro_specs}. The random seed for all stochastic processes was set to 42.
	The hyperparameters listed in Table~\ref{tab:training_hyperparams} were selected based on a series of sweeps and found to be robust for this problem class; a detailed sensitivity analysis is left for future work.
	
	\begin{table}[htbp]
		\centering
		\caption{Hardware and Software Specifications.}
		\label{tab:repro_specs}
		\small
		\renewcommand{\arraystretch}{1.1}
		\begin{tabular}{@{}ll@{}}
			\toprule
			\textbf{Component} & \textbf{Specification} \\
			\midrule
			CPU & AMD Ryzen 7 \\
			GPU & NVIDIA GeForce RTX 4070 (8 GB VRAM) \\
			RAM & 40 GB \\
			OS & Windows 11 \\
			Python & 3.12 \\
			Key Libraries & PyTorch, PyTorch Geometric, PyPower, Numba \\
			\bottomrule
		\end{tabular}
	\end{table}
	
	\section*{Declaration of generative AI and AI-assisted technologies in the writing process}
	During the preparation of this work the author used Gemini 2.5 Pro in order to improve the language and clarity of the manuscript. After using this tool/service, the author reviewed and edited the content as needed and takes full responsibility for the content of the published article.

	\FloatBarrier

	\begin{IEEEbiography}[{\includegraphics[width=1in,height=1.25in,clip,keepaspectratio]{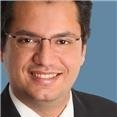}}]{Mohamed Shamseldein}
		(S'09--M'13--SM'19) received the B.Sc. and M.Sc. degrees in electrical engineering from Ain Shams University, Cairo, Egypt, and the Ph.D. degree in electrical and computer engineering from the University of Waterloo, Waterloo, ON, Canada.
		
		He is currently an Assistant Professor with the Department of Electrical Power and Machines, Faculty of Engineering, Ain Shams University, Cairo, Egypt. His research interests include power system analysis, renewable integration, and the application of AI/ML for grid analysis. Dr. Shamseldein is a licensed Professional Engineer (P.Eng.) in the province of Ontario.
	\end{IEEEbiography}
	
	\vfill
	
\end{document}